\documentclass[12pt]{iopart}

\usepackage{iopams,graphicx,amssymb}
\def\eps{\varepsilon}
\def\Dm{\widetilde{\cal D}_{\mu}}

\def\x{{\bf x}}
\def\k{{\bf k}}
\def\q{{\bf q}}
\def\r{{\bf r}}
\begin{document}
\title[Effects of mixing and stirring on the critical behaviour]{Effects of mixing and stirring on the critical behaviour}

\author{N V~Antonov,$^1$ Michal Hnatich$^{2,3,4}$ and Juha~Honkonen$^{5,6}$}
\address{$^1$Department of Theoretical Physics, St.~Petersburg University,
Uljanovskaja 1, \\ St.~Petersburg, Petrodvorez, 198504 Russia}
\address{$^{2}$ Institute of Experimental Physics, Slovak Academy
of Sciences, Watsonova 47, 04011 Ko\v{s}ice, Slovakia}
\address{
$^{3}$ Faculty of Civil Engineering, Technical University,
Vysoko\v{s}kolsk\'a 4, 043 53 Ko\v{s}ice, Slovakia}
\address{$^{4}$ N.\,N.~Bogoliubov Laboratory of Theoretical Physics,
Joint Institute for Nuclear Research,
141980 Dubna, Moscow Region, Russia}
\address{$^5$Division of Theoretical Physics, Department~of~Physics,
 FIN-00014 University~of~Helsinki, Finland}
\address{$^6$Department of Military Technology, National Defence College,
FIN-00861 Helsinki, Finland}
\ead{juha.honkonen@helsinki.fi}

\begin{abstract}
Stochastic dynamics of a nonconserved scalar order parameter near its
critical point, subject to random stirring and mixing, is studied using
the field theoretic renormalization group. The stirring and mixing are
modelled by a random external Gaussian noise with the correlation
function $\propto\delta(t-t') k^{4-d-y}$ and the divergence-free
(due to incompressibility) velocity field, governed
by the stochastic Navier--Stokes equation with a random Gaussian force
with the correlation function $\propto\delta(t-t') k^{4-d-y'}$. Depending
on the relations between the exponents $y$ and $y'$ and the space
dimensionality $d$, the model reveals several types of scaling regimes.
Some of them are well known (model A of equilibrium critical dynamics and
linear passive scalar field advected by a random turbulent flow), but
there are three new nonequilibrium regimes (universality classes) associated
with new nontrivial fixed points of the renormalization group equations.
The corresponding critical dimensions are calculated in the  two-loop
approximation (second order of the triple expansion in $y$, $y'$ and
$\varepsilon=4-d$).
\end{abstract}

\pacs{ 64.75.+g, 05.10.Cc, 64.60.Ht, 05.40$-$a} \maketitle
\normalsize

\section{Introduction} \label{sec:Intro}

Over the past three decades, increasing attention has been attracted by the
dynamics of phase ordering --- the growth of order through domain coarsening
(spinodal decomposition), when a system (e.g. a ferromagnet or a binary
alloy) is quenched from its high-temperature homogeneous phase into the
low-temperature multi-phase coexistence region; see
\cite{Adv}--\cite{Angelo} and references therein.

Much interest was focused on the late stages of the coarsening
process, when some kind of a self-similar (scaling) regime
develops with apparently universal exponents --- the features
normally associated with the critical behaviour. That regime is by
now rather well understood; see Ref. \cite{Adv} and the reviews
cited the. Phenomenological approaches, renormalization group (RG)
techniques, exactly soluble models and numerical simulations show
that the characteristic domain size increases as a power of time,
$L(t) \sim t^{\alpha}$, where the growth exponent $\alpha$ depends
on the global characteristics of the system (conserving or
nonconserving dynamics, scalar or vector order parameter,
dimensionality of space) but not on its detailed structure (like
the values of the coupling constants). Therefore, in recent years
attention has been directed to systems subjected to external
stirring, like binary mixtures under imposed shear flow or other
kinds of deterministic or random (e.g. turbulent) velocity fields;
see \cite{Aron}--\cite{Chan} and references therein.

Numerical experiments and theoretical analysis (e.g. the linear stability
analysis of the corresponding dynamic equations) of binary alloys subjected
to statistically isotropic and homogeneous random velocity ensembles of very
different kinds also suggest that, at least close to the critical point and
under vigorous stirring, the domain growth is ``arrested'' and a new
dynamical nonequilibrium steady state emerges, which is characterized
by a continuous formation and rupture of finite-size domains
\cite{Aron,Stirring,Stirring2,Angelo}.

Emergence of the nonequilibrium steady states appears rather a generic and
robust phenomenon, being observed in two-dimensional numerical simulations
for passive \cite{Stirring,Stirring2,Angelo} and active \cite{Angelo} order
parameters subjected to a random Gaussian velocity field with finite
correlation length and time \cite{Stirring} and various kinds of regular and
chaotic cellular flows \cite{Stirring2,Angelo}. The questions which
naturally arise within this context, and which will be addressed in the
present paper, are the following: Do those steady states reveal some kinds
of self-similar behaviour? Do the corresponding correlation and structure
functions exhibit power laws? If yes, do those states belong to the
universality classes known for the models of equilibrium critical dynamics
\cite{HH,FM}, or do they represent new types of scaling behaviour?
Are there any crossover dimensions for the new scaling regimes? Is it
possible to establish the existence of these scaling regimes on the basis
of microscopic models, and to calculate the corresponding exponents in
consistent approximations or, better, within regular perturbation
expansions? To what extent this behaviour is universal? What are the
parameters the scaling dimensions depend on?

We will consider the dynamics of a scalar (one-component) passive
(no feedback on the velocity field) nonconserving order parameter
$\varphi(x) \equiv \varphi(t,\x)$ governed by the stochastic equation
\begin{eqnarray}
\sigma_{0} \nabla_{t} \varphi = \partial^{2} \varphi - V'(\varphi) + f,
\qquad \nabla_{t} = \partial_{t} + v_{i} \partial_{i},
\label{eq0}
\end{eqnarray}
where $\sigma_{0}>0$ is the reciprocal of the kinetic coefficient and the
potential $V(\varphi)$ will be chosen as in the well-known models of
critical dynamics \cite{HH,FM,Book3,Zinn}. However,
in contrast to the latter,
the stirring noise $f(x)$ and the velocity $v_{i}(x)$ are not chosen
such that the steady state of the system is in equilibrium, or, in other
words, its equal-time correlation functions are not described by the
Landau--Ginzburg Hamiltonian. Namely, the transverse (divergence-free,
due to the incompressibility condition $\partial_i v_i=0$) velocity field
satisfies the Navier--Stokes equation with a random driving force
\begin{equation}
\nabla _t v_i=\nu _0\partial^{2} v _i-\partial _i {\cal P}+f_i,
\label{1.1}
\end{equation}
where ${\cal P}$ and $f_i$ are the pressure and the transverse random force
per unit mass (all these quantities depend on $x$).

The random sources $f(x)$ and $f_{i}(x)$ maintain the steady state of the
system and model the effects of external stirring and/or shaking and initial
and/or boundary conditions. The use of such random stirring terms
is a commonplace in the statistical theory of turbulence
\cite{FNS,DeDom,Frisch,Red} and other nonequilibrium phenomena
\cite{FNS,KPZ}: it allows one to do away with the details of the geometry
of the system and to consider a homogeneous and isotropic problem in the
infinite space. Let us specify their statistical properties.

In models of equilibrium critical dynamics the form of such correlators
for Langevin equations (like e.g. (\ref{1.1}) without the
velocity) is uniquely determined by the requirement that the dynamics and
statics be mutually consistent (that is, the equal-time correlations
of the dynamical problem be given by the Landau--Ginzburg Hamiltonian);
see \cite{HH,FM,Book3}. Such arguments do not apply to our non-equilibrium
model. Like in the RG theory of turbulence, the correlators will
be chosen on the basis of both physical and technical arguments.

Consider for definiteness the correlation function
\begin{eqnarray}
D_{\varphi}(x,x')\equiv \langle f(x)f(x') \rangle = \delta(t-t')\, D(r),
\qquad r=|\x-\x'|,
\label{stepen}
\end{eqnarray}
of the source field of the stochastic equation (\ref{eq0}). The
function $D(r)$ depends only on $r=|\x-\x'|$, its Fourier
transform being $D(k)$. The physical arguments are that the noises
model the injection of energy to the system owing to interaction
with the large-scale stirring. Thus for realistic case the
dominant contribution to the correlators $D(k)$ must come from
small momenta $k \sim m$, where $m=1/L$ is the reciprocal of the
integral (external) scale $L$ (the size of the system or a
stirring device). Idealized injection by infinitely large modes
corresponds to $D(k) \propto \delta({\bf k})$. On the other hand,
for the use of the standard RG technique it is important that the
function $D(k)$ have a power-law behaviour at large $k$. This
condition is satisfied if $D(k)$ is chosen in the form
\cite{DeDom} $D(k)=D_0\,h(m/k)\,k^{4-d-y}$, where $D_0>0$ is an
amplitude factor, $d$ is the space dimension and the exponent $y$
plays the part analogous to that played by $4-d$ in the RG theory
of critical behaviour. The function $h(m/k)$ with $h(0)=1$
provides the IR regularization. Its specific form is unessential;
we will use the sharp cutoff $h(m/k)= \theta(m-k)$ with the
Heaviside step function to simplify the practical calculation (in
the calculations in the spirit of dimensional regularization one
could simply set $h=1$).

The large-scale forcing is reproduced in the limit $y\to4$, as follows from
the well-known power-law representation of the $d$-dimensional $\delta$
function,
\begin{eqnarray}
\delta({\bf k}) &=& \lim_{y\to 4}\, \frac{1}{(2\pi)^{d}} \int
d{\bf x} \, (\Lambda x)^{y-4} \, \exp [{\rm i} ({\bf k}\cdot{\bf x})] =
\nonumber \\
 &=& S_{d}^{-1} k^{-d} \lim_{y\to 4} \left[ (4-y)
(k/\Lambda)^{4-y} \right],
\label{powerdelta}
\end{eqnarray}
where $S_d$ is the surface area of the unit sphere in $d$-dimensional space
(see (\ref{26}) in section~\ref{sec:velocity}) and $\Lambda$ has the
dimension of a momentum. Representation (\ref{powerdelta}) also specifies
the appropriate choice of the amplitude $D_{0}$ at $y\to4$. More detailed
discussion of this issue can be found in section~6.3 of book \cite{Book3}.

Following these ideas, the sources $f(x)$ and $f_{i}(x)$ will be taken
Gaussian, white in time (this is dictated by the principle of maximum
entropy \cite{maximum}) and with power-law spectra,
$\langle ff \rangle \propto \delta(t-t') k^{4-d-y}$ for the scalar noise
and $\langle f_{i}f_{j} \rangle \propto \delta(t-t') P_{ij}(\k) k^{4-d-y'}$
for the vector one, where $\k$ is the momentum (wave vector), $k=|\k|$ is the
wave number, $d$ is the space dimensionality, $P_{ij}({\bf k})=\delta _{ij}
-k_i k_j/k^2$ is the transverse projector, and $y$ and $y'$ are arbitrary
parameters. The large-scale forcing corresponds to the limits
$y$, $y'\to 4$.

The time decorrelation of the random force guarantees that the
full stochastic problem (\ref{eq0}), (\ref{1.1}) is Galilean
invariant for all values of the model parameters, including $D_0$
and $d$. As a consequence, the ordinary perturbation theory for
the model (the expansion in the nonlinearities) is manifestly Galilean
covariant: all the exact relations between the correlation
functions imposed by the Galilean symmetry (Ward identities) are
satisfied order by order. The renormalization procedure does not
violate the Galilean symmetry, so that the improved perturbation
expansion, obtained with the aid of RG, also remains covariant.

In a wider context, the model (\ref{eq0}), (\ref{1.1}) can be
interesting as a model system for studying generic nonequilibrium
dynamical features. Recently, significant progress has been
achieved in classifying large-scale, long-distance scaling
behaviour of such phenomena, including driven diffusive systems,
diffusion-limited reactions, growth, ageing and percolation
processes, and so on; see e.g. \cite{driven} and references
therein. Being analytically tractable, our model can serve as a
good testing ground in studying such scaling regimes and their
universality. Similar (but nonstationary) models also arise in
stochastic inflationary models designed to describe the structure
formation in cosmology; see \cite{Inflat}.

We will apply to the stochastic problem (\ref{eq0}) the field theoretic
renormalization group (RG), which proved to be extremely useful in describing
equilibrium critical phenomena, including their kinetic properties
\cite{HH,FM,Book3,Zinn}. In the RG framework, long-wavelength scaling
regimes are associated with infrared (IR) attractive fixed points of the
corresponding RG equations. Earlier, the RG approach was applied to the
problem of phase separation and domain growth in a number of studies
\cite{Adv,Kine1,Kine2,MCRG,BrayRG}. By contrast with critical phenomena,
however, the application of the RG to the growth problems suffers from the
lack of an (obvious) small parameter (analogous to $\eps=4-d$ for the
Landau--Ginzburg model and corresponding dynamical models), the problem also
encountered for the stochastic Burgers and Kardar--Parisi--Zhang models
\cite{FNS,KPZ}. To circumvent this obstacle, in
\cite{Adv,Kine1,Kine2,MCRG,BrayRG}, the RG was used in the form of block-spin
transformations performed using numerical Monte Carlo simulations, proposed
earlier in \cite{Ma}. Another possibility, explored in \cite{BrayRG}
(see also discussion in the review paper \cite{Adv}), was
to assume the existence of the RG symmetry and an appropriate strong-coupling
fixed point, and then to use specific features of the conserved dynamics
(absence of renormalization of the transport coefficient, well known for the
model B of equilibrium dynamics \cite{HH,FM,Book3}) to derive some exact
relations between the critical exponents. To complete that analysis, however,
one should take some exponents from the experiment or derive them using
additional phenomenological considerations \cite{Adv,BrayRG}.

The plan of the paper is as follows. We begin with the analysis of the
model without velocity, which appears nontrivial and reveals a new type of
scaling behaviour. In section~\ref{sec:FT} we present the field theoretic
formulation of the model and its renormalization. After an appropriate
extension, the model becomes multiplicatively renormalizable, and the
differential RG equations can be derived in a standard fashion
(section~\ref{sec:RGE}). The fixed points and their regions of IR stability
are analyzed in section~\ref{sec:Fixed}. It is shown that a systematic
perturbation
expansion in the two parameters, $y$ and $\eps=4-d$, can be constructed for
the coordinates of the fixed points and critical dimensions, with the
additional assumption that $y-\eps=O(\eps^{2})$. One of the two nontrivial
fixed points corresponds to the well known model A of equilibrium critical
dynamics, while the other represents a new nonequilibrium universality
class; the corresponding critical dimensions are calculated in the two-loop
approximation (section~\ref{sec:scaling}). In section~\ref{sec:velocity}, the full
model with the velocity field, governed by the stochastic Navier--Stokes
equation, is studied. Two additional nonequilibrium scaling regimes
(universality classes) are identified; the corresponding dimensions are
found to the second order of the triple expansion in $y$, $y'$ and
$\eps=4-d$. Section~\ref{sec:Conc} is reserved for a brief conclusion. Some
interesting details of the two-loop calculation are given in \ref{sec:Zs}.

The field-theoretic renormalization group was earlier applied to the problem
of the effects of turbulence on the critical behaviour of binary mixtures
in \cite{Satten}. The model studied in that work was less realistic
than the present one in two respects: turbulence was modelled by a Gaussian
time-decorrelated statistical ensemble and the noise was taken to be purely
thermal. On the other hand, our model is less realistic in the sense that
the order parameter here is not conserved. Nevertheless, the main qualitative
conclusion drawn from the two cases is the same: the instability of the
equilibrium fixed point and the existence of a new non-equilibrium critical
regime was established. Thus we may conclude that such a phenomenon appears
quite robust and insensitive to the details of the model.

\section{The model without convection: Field theoretic formulation
and renormalization} \label{sec:FT}

It is instructive to begin the analysis with the model with no convection
term in (\ref{eq0}), which already exhibits a nonequilibrium scaling regime
and involves some interesting formal subtleties. The dynamical equation
for the order parameter $\varphi(x) \equiv \varphi(t,\x)$ then becomes
\begin{eqnarray}
\sigma_{0}\partial_{t} \varphi = \partial^{2} \varphi -V'(\varphi) + f.
\label{eq1}
\end{eqnarray}
Correlator of the random noise $f(x)$ will be taken in the form
\begin{eqnarray}
D_{\varphi}(x,x')\equiv \langle f(x)f(x') \rangle = \delta(t-t')\, D(r),
\qquad r=|\x-\x'|,
\label{forceD}
\end{eqnarray}
with some function $D(r)$ depending only on $r=|\x-\x'|$. The
choice $D(r)= 2\sigma_{0} \delta(\x-\x')$ corresponds to the
well-known model A of critical dynamics, which describes kinetic
properties of the equilibrium critical state
\cite{HH,FM,Book3,Zinn}; the probability distribution function of
its equal-time correlators is then given by $\exp\left(-H(\varphi)
\right)$ where $H(\varphi)= -
\varphi\partial^{2}\varphi+V(\varphi)$ with the implied
integration over $\x$ is the Hamiltonian for the time-independent
field $\varphi(\x)$.

We assume that the model is near its critical point, and, in the spirit of
the Landau theory, retain in $V(\varphi)$ only the first terms of the Taylor
expansion: $V(\varphi)= \tau_{0} \varphi^{2}/2 + \lambda_{0} \varphi^{4}/24$,
where $\tau_{0}$ is the deviation of the temperature from its critical
value. The
function $D(r)$ in (\ref{forceD}), however, will be chosen in the power-like
form $D(r) \propto r^{-4+y}$, which in the momentum representation gives
\begin{eqnarray}
D(k) =D_{0} k^{4-d-y},
\label{force}
\end{eqnarray}
where $k=|\k|$ is the wave number, $y$ is an arbitrary parameter and
$D_{0}>0$ an amplitude factor. The IR cutoff at $k=m$ is implied.

According to the general theorem (see e.g. \cite{Book3,Zinn}), stochastic
problem (\ref{eq1}), (\ref{forceD}) is equivalent to the field theoretic
model of the doubled set of fields $\Phi\equiv\{\varphi', \varphi\}$ with
action functional
\begin{eqnarray}
S(\Phi) =  \varphi' D_{\varphi} \varphi' /2 + \varphi' \left\{-
\sigma_{0}\partial_{t} \varphi + \partial^{2} \varphi - \tau_{0} \varphi
- \lambda_{0} \varphi^{3} /6 \right\},
\label{MSR}
\end{eqnarray}
with $D_{\varphi}$ from (\ref{forceD}) and implied integrations over the
argument $x=\{t,{\bf x}\}$. Formulation (\ref{MSR}) means that statistical
averages of random quantities in the original stochastic problem can be
represented as functional averages with the weight $\exp S(\Phi)$. The model
(\ref{MSR}) corresponds to a standard Feynman diagrammatic technique with two
bare propagators (lines in the diagrams) $\langle\varphi\varphi\rangle_{0}$
and $\langle \varphi \varphi' \rangle_{0}$ (their explicit form is given in
\ref{sec:Zs}) and the vertex $\varphi'\varphi^{3}$.

The analysis of ultraviolet (UV) divergences is based on the analysis of
canonical dimensions. Dynamical models of the type (\ref{MSR}), in contrast
to static models, have two scales, i.e., the canonical dimension of some
quantity $F$ (a field or a parameter in the action functional) is described
by two numbers, the momentum dimension $d_{F}^{k}$ and the frequency
dimension $d_{F}^{\omega}$. They are determined such that
$[F] \sim [L]^{-d_{F}^{k}} [T]^{-d_{F}^{\omega}}$, where $L$ is the
length scale and $T$ is the time scale. The dimensions are found
from the obvious normalization conditions $d_k^k=-d_{\bf x}^k=1$,
$d_k^{\omega}=d_{\bf x}^{\omega }=0$, $d_{\omega }^k=d_t^k=0$,
$d_{\omega }^{\omega }=-d_t^{\omega }=1$, and from the requirement
that each term of the action functional be dimensionless (with
respect to the momentum and frequency dimensions separately).
Then, based on $d_{F}^{k}$ and $d_{F}^{\omega}$, one can introduce the total
canonical dimension $d_{F}=d_{F}^{k}+2d_{F}^{\omega}$ (in the free theory,
$\partial_{t}\propto\partial^{2}$), which plays in the theory of
renormalization of dynamical models the same role as the conventional
(momentum) dimension does in static problems, see e.g. \cite{Book3}.
The resulting canonical dimensions are given in table~1,
including the dimensions of the parameters which will appear later on
(renormalized parameters and the others). It is easily checked that the
role of the coupling constant (expansion parameter in the ordinary
perturbation theory) in model (\ref{MSR}) with correlator (\ref{forceD})
is played by the combination $\lambda_{0} D_{0} /\sigma_{0}$. From table~1
it follows that this constant has the dimension $\Lambda^{y}$
with some momentum scale $\Lambda$. Thus the case $y<0$ corresponds to
the Gaussian IR behaviour (perturbation theory works in the IR range),
$y=0$ is the logarithmic value, and for $y\ge0$ the RG summation is needed.
The UV divergences have the form of the poles in $y$ in the correlation
functions of the fields $\Phi \equiv\{\varphi', \varphi\}$.

The total canonical dimension of an arbitrary 1-irreducible correlation
function $\Gamma = \langle\Phi\cdots\Phi\rangle $ is given by the relation
$d_{\Gamma }=d_{\Gamma }^k+2d_{\Gamma }^{\omega }= d+2-N_{\Phi }d_{\Phi}$,
where $N_{\Phi}=\{N_{\varphi},N_{\varphi'}\}$ are the numbers of corresponding
fields entering into the function $\Gamma$, and the summation over all types
of the fields is implied. The total dimension $d_{\Gamma}$ in the logarithmic
theory (that is, at $y=0$) is the formal index of the UV divergence.
Superficial UV divergences, whose removal requires counterterms, can be
present only in those functions $\Gamma$ for which $d_{\Gamma}$ is a
non-negative integer. Straightforward analysis shows that for all $d>4$,
superficial UV divergences can be present only in the 1-irreducible
functions $\langle \varphi' \varphi \rangle$ with the counterterms
$\varphi' \partial _{t}\varphi$, $\varphi' \partial^{2} \varphi$,
$\tau_{0} \varphi' \varphi$, and $\langle \varphi' \varphi^{3} \rangle$
with the counterterm  $\varphi' \varphi^{3}$.  Such terms are present in the
action (\ref{MSR}), so that the model is multiplicatively renormalizable.

However, for $d\le 4$ a new divergence appears in the function
$\langle \varphi' \varphi' \rangle$ (for $d=4$ and small $y$, the
noise correlator becomes almost polynomial in $\k$ --- namely constant ---
and local in $\x$ representation). It would be erroneous to try to
eliminate this divergence by renormalizing the nonlocal noise term
(as thoroughly discussed in Ref. \cite{local} for the stochastic
Navier--Stokes equation). We therefore must add the local counterterm of the
form $\varphi'\varphi'$. So we are forced to consider the variable space
dimension and, for $d\le 4$, to extend the original model (to include the
local term in the noise correlator from the very beginning) in order to have
multiplicative renormalizability. Finally we arrive at the extended model
\begin{eqnarray}
\sigma_{0}\partial_{t} \varphi =\partial^{2} \varphi - \tau_{0} \varphi
- \lambda_{0} \varphi^{3} /6 + f,
\label{model1} \\
\langle f(x)f(x') \rangle = 2\sigma_{0}\, \delta(t-t') \left\{
w_{0}\, k^{4-d-y} + 1 \right\}
\label{model}
\end{eqnarray}
which has become multiplicatively renormalizable for $d<4$; the
special case $w_{0}=0$ gives the model A (which is
multiplicatively renormalizable in itself). Interpretation of the
additional local term in (\ref{model}) can be twofold. On the one
hand, the fact that it is generated by the renormalization
procedure means that it is not forbidden by dimensionality or
symmetry considerations and, therefore, it is natural to include
it in the model from the very beginning. In the language of the
Wilsonian RG, this means that such term necessarily arises in the
effective model for the properly smoothed (coarse-grained) field;
it becomes IR relevant for $d<4$, where it affects the critical
behaviour and cannot be neglected. On the other hand, one can
insist on studying the original model with a purely power-law
correlation function. Then the extension of the model is only
needed to ensure the multiplicative renormalizability and to
derive the RG equations; the latter should be solved with the
special initial data that correspond to the power-law correlator.
Since the IR attractive fixed point of the RG equations is unique
for any given choice of the parameters $\eps$ and $y$ (see section
\ref{sec:Fixed}), the resulting IR behaviour is the same as for
the case of the general correlation function (\ref{model}) with
the inclusion of the local term.

By dimension the couplings $\lambda_{0}\sim \Lambda^{\eps}$ and
$w_{0}\sim \Lambda^{y-\eps}$ with $\eps = 4-d$, so that we expect the
double RG expansion in $y$ in $\eps$ instead of a single expansion in $y$
(as for $d>4$) or in $\eps$ (as for the model A). However, as we will see,
the real situation appears slightly more complicated.

The renormalized action is
\begin{eqnarray}
S_{R}(\varphi,\varphi') =  \sigma \varphi' \left\{
w\mu^{y-\eps} k^{\eps-y} + Z_{1}  \right\} \varphi'
\nonumber \\ + \varphi' \left\{
- Z_{2}\sigma\partial_{t} \varphi + Z_{3}\partial^{2} \varphi
- \tau Z_{5} \varphi - Z_{4}\lambda \mu^{\eps}\varphi^{3} /6 \right\},
\label{MSRR}
\end{eqnarray}
which is equivalent to the multiplicative renormalization of the fields
$\varphi \to Z_{\varphi} \varphi$, $\varphi' \to Z_{\varphi}' \varphi'$
and parameters
\begin{eqnarray}
\sigma_{0}=Z_{\sigma}\sigma, \quad \tau_{0}=Z_{\tau}\tau, \quad
u_{0}= \mu^{\eps} Z_{u} u, \quad w_{0}= \mu^{y-\eps} Z_{w} w,
\label{reno}
\end{eqnarray}
where we introduced the new coupling constant $u = \lambda/16\pi^2$
(the coefficients in RG functions become slightly simpler) and $\mu$
is the reference mass -- additional parameter of the renormalized theory.

The renormalization constants $Z_{i}=Z_{i}(\eps,y,g,w)$ capture the
divergences at $\eps,y\to 0$, so that the correlation functions of the
renormalized model (\ref{MSRR}) have finite limits for $\eps,y = 0$
(when expressed in renormalized parameters $u$, $w$, $\tau$ and $\mu$).

The relations between the $Z$'s in (\ref{MSRR}) and (\ref{reno}) have
the forms
\begin{eqnarray}
Z_{\sigma} Z_{\varphi'}^{2}= Z_{1}, \quad
Z_{\sigma} Z_{\varphi'}^{2} Z_{w}=1  , \quad
Z_{\sigma} Z_{\varphi'} Z_{\varphi} = Z_{2} ,     \nonumber \\
Z_{\varphi}Z_{\varphi'}=Z_{3}, \quad
Z_{\tau}Z_{\varphi}Z_{\varphi'}=Z_{5}, \quad
Z_{u} Z_{\varphi}^{3}Z_{\varphi'}=Z_{4}.
\label{Z}
\end{eqnarray}

\begin{figure}
\begin{center}
\includegraphics[width=10cm]{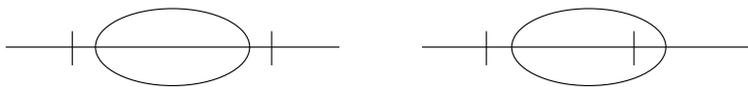}
\caption{\label{fig1} Two-loop graphs giving rise to
renormalization constants $Z_{1}$--$Z_{3}$. The simple lines
correspond to the bare propagator $\langle \varphi \varphi
\rangle_{0} =  2\sigma\, \left\{ w\, (k/\mu)^{y-\eps} + 1
\right\}/ (\omega^{2} \sigma^{2}+ k^{4})$ in frequency-momentum
representation, whereas the lines with a slash to the propagator $
\langle \varphi \varphi' \rangle_{0} = (-{\rm
i}\sigma\omega+k^{2})^{-1} $, the slash indicating the variables
of the the field $\varphi'$. The vertex factor is
$\lambda\mu^{\eps}$.}
\end{center}
\end{figure}

The renormalization constants $Z_{1}$--$Z_{5}$ are calculated according to
standard rules from the perturbation theory; then the constants in
(\ref{reno}) are easily found using the relations (\ref{Z}).
The expansion parameter in the $Z$'s is $u$, while
the dependence on the second coupling constant $w$ should be calculated
exactly in each order of the expansion in $u$. Like for the model A, the
first nontrivial contributions to the constants $Z_{4,5}$ are determined by
one-loop Feynman graphs, so that $Z_{4,5}=1+O(u)$. The leading
contributions to the constants $Z_{1}$--$Z_{3}$ are determined by
two-loop (``watermelon'') graphs depicted in figure \ref{fig1}, so that $Z_{1,2,3}=1+O(u^{2})$. These
details will be important in the analysis of the fixed points of the RG
equations. The two-loop calculation of the renormalization constants in our
model is illustrated by the example of $Z_{1}$ in \ref{sec:Zs}.

\section{RG equations and RG functions} \label{sec:RGE}

Let us recall an elementary derivation of the RG equations; see
\cite{Book3}. The RG equations are written for the renormalized
correlation functions $G_{R} =\langle \Phi\cdots\Phi\rangle_{R}$, which
differ from the original (unrenormalized) ones
$G =\langle \Phi\cdots\Phi\rangle$ only by normalization and choice of
parameters, and therefore can be equally used for analyzing the critical
behaviour. The relation $S_{R} (\Phi,e,\mu) = S (\Phi,e_{0})$ between the
functionals (\ref{MSR}) and (\ref{MSRR}) results in the relations
$G(e_{0},\dots) = Z_{\varphi}^{N_{\varphi}} Z_{\varphi'}^{N_{\varphi'}}
G_{R}(e,\mu,\dots)$ between the correlation functions. Here, as usual,
$N_{\varphi}$ and $N_{\varphi'}$
are the numbers of corresponding fields entering into $\Gamma$; $e_{0}=\{
\sigma_{0}, \tau_{0}, w_{0}, \lambda_{0}\}$ is the full set of bare parameters and
$e=\{ \sigma, \tau, w, \lambda\propto u\}$ are their renormalized analogs;
the dots stand for the other arguments (times, coordinates, momenta etc).
We use $\widetilde{\cal D}_{\mu}$ to denote the differential operation
$\mu\partial_{\mu}$ for fixed $e_{0}$ and operate on both sides of this
equation with it. This gives the basic RG differential equation:
\begin{equation}
\left\{ {\cal D}_{RG} + N_{\varphi}\gamma_{\varphi} + N_{\varphi'}
\gamma_{\varphi'}
\right\} \,G^{R}(e,\mu,\dots) = 0,
\label{RG1}
\end{equation}
where ${\cal D}_{RG}$ is the operation $\widetilde{\cal D}_{\mu}$
expressed in the renormalized variables:
\begin{equation}
{\cal D}_{RG}\equiv {\cal D}_{\mu} + \beta_{u}\partial_{u} +
\beta_{w}\partial_{w} - \gamma_{\sigma}{\cal D}_{\sigma}
- \gamma_{\tau}{\cal D}_{\tau}.
\label{RG2}
\end{equation}
In equation (\ref{RG2}), we have written ${\cal D}_{x}\equiv x\partial_{x}$ for
any variable $x$, the RG anomalous dimensions $\gamma$ are defined as
\begin{equation}
\gamma_{F}\equiv \Dm \ln Z_{F} \quad {\rm for\ any\ quantity} \ F,
\label{RGF1}
\end{equation}
and the $\beta$ functions for the two dimensionless couplings $u$ and
$w$ are
\refstepcounter{equation}
\label{Betas}
\addtocounter{equation}{-1}
\numparts
\begin{eqnarray}
\beta_{u} \equiv \tilde {\cal D}_{\mu} u
&=& u [-\eps -\gamma_{u}],
\label{betau} \\
\beta_{w} \equiv \tilde {\cal D}_{\mu} w
&=& w [-y+\eps -\gamma_{w}],
\label{betaw}
\end{eqnarray}
\endnumparts
where the last equalities come from the definitions and relations
(\ref{reno}).

The anomalous dimensions $\gamma_{1}$--$\gamma_{5}$ are found from the known
constants $Z_{1}$--$Z_{5}$ (see \ref{sec:Zs} for $\gamma_{1}$). Then the
relations (\ref{Z}) give
\begin{eqnarray}
\gamma_{\sigma}+ 2\gamma_{\varphi'}= \gamma_{1}, \quad
\gamma_{\sigma}+ 2\gamma_{\varphi'}+ \gamma_{w} = 0 , \quad
\gamma_{\sigma}+ \gamma_{\varphi'}+ \gamma_{\varphi} = \gamma_{2},
\nonumber \\
\gamma_{\varphi}+\gamma_{\varphi'}=\gamma_{3}, \quad
\gamma_{\tau}+\gamma_{\varphi}+\gamma_{\varphi'}=\gamma_{5}, \quad
\gamma_{u}+3 \gamma_{\varphi}+\gamma_{\varphi'}=\gamma_{4}.
\label{Z1}
\end{eqnarray}
Resolving these relations gives
\begin{eqnarray}
\gamma_{\sigma}=\gamma_{2}-\gamma_{3}, \quad
\gamma_{\tau}=\gamma_{5}-\gamma_{3},   \quad
\gamma_{u}= \gamma_{1}-\gamma_{2}-2\gamma_{3}+\gamma_{4},
\nonumber \\
\gamma_{w}= -\gamma_{1} , \quad
2\gamma_{\varphi}= \gamma_{3}+\gamma_{2}-\gamma_{1}, \quad
2\gamma_{\varphi'}= \gamma_{1}+\gamma_{3}-\gamma_{2},
\label{sol}
\end{eqnarray}
and for the $\beta$ functions (\ref{Betas}) one obtains:
\begin{eqnarray}
\beta_{u} = u [-\eps - \gamma_{1} + \gamma_{2} +2\gamma_{3} - \gamma_{4}],
\qquad
\beta_{w} = w [-y+\eps +\gamma_{1}].
\label{betas}
\end{eqnarray}

\section{Fixed points and scaling regimes}  \label {sec:Fixed}

It is well known that possible scaling regimes of a renormalizable
model are associated with the IR attractive fixed points of the
corresponding RG equations. In our model, the coordinates
$u_{*}$, $w_{*}$ of the fixed points are found from the equations
\begin{equation}
\beta_{u} (u_{*},w_{*})=0, \quad \beta_{w} (u_{*},w_{*})=0
\label{points}
\end{equation}
with the beta functions given in (\ref{betas}).
The type of a fixed point is determined by the matrix
$\Omega=\{\Omega_{ij}=\partial\beta_{i}/\partial g_{j}\}$,
where $\beta_{i}$ denotes the full set of the beta functions and
$g_{j}= \{u,w\}$ is the full set of couplings. For IR stable fixed points
the matrix $\Omega$ is positive, i.e., the real parts of all its
eigenvalues are positive.

From the forms of the renormalization constants (see the remark in the end
of section~\ref{sec:FT}) it follows that $\gamma_{4,5}=O(u)$ while
$\gamma_{1,2,3,4}=O(u^{2})$. Therefore only $\gamma_{4}$ gives the leading
contribution to the function $\beta_{u}$ in (\ref{betas}). The
actual calculation gives (see \ref{sec:Zs})
\begin{equation}
\gamma_{1}=bu^{2}(1+w)^{3}+O(u^{3}) , \qquad \gamma_{4}=- au(1+w)+O(u^{2})
\label{gammas}
\end{equation}
with $a=3$ and $b=\ln (4/3)$. Thus the leading-order expressions for the
$\beta$ functions are
\begin{eqnarray}
\beta_{u} &=& u [-\eps + au (1+w) +O(u^{2}) ],
\nonumber \\
\beta_{w} &=& w [-y+\eps + bu^{2}(1+w)^{3}  +O(u^{3}) ].
\label{betas2}
\end{eqnarray}
From (\ref{betas2}) one immediately finds the local Gaussian fixed point
$u_{*}=w_{*}=0$, which is IR attractive for $\eps<0$, $\eps>y$. The case
$u_{*}=0$, $w_{*}\ne0$ appears more subtle. Substituting $u_{*}=0$ into
$\beta_{w}$ gives $\beta_{w} = w (-y+\eps)$, which suggests that for
$\eps<0$, $\eps<y$, the IR attractive point has $w_{*}=\infty$. Then,
however, the ambiguity is encountered in $u^{2}(1+w)^{3}$ and the
higher-order terms in $\beta_{w}$. To resolve it, one should recall that
for $\eps<0$, no local counterterm to the noise correlator is in fact
needed, and we can consider the original problem with a purely nonlocal
correlator (\ref{force}). For such a model, one easily finds that $y<0$
corresponds to the Gaussian fixed point with finite amplitude of the
noise correlator (in fact, for the purely nonlocal case in can simply be
scaled out from the action).
The case $y>0$, $\eps<0$ requires a bit more elaborate
analysis, which shows that the large-scale behaviour also in this region
is that of the Gaussian model with the nonlocal correlator. Thus,
we finally may conclude that the region
$\eps<0$, $\eps<y$ corresponds to the Gaussian fixed point with
$u_{*}=0$ and purely nonlocal noise correlator.

In models with two regulators like $\eps$ and $y$, it is usually implied that
they are of the same order, $y=O(\eps)$, and the coordinates of the fixed
points and the values of anomalous dimensions at those points are sought in
the form of double series in $\eps$ and $y$; see e.g.~\cite{local}. No
such solution, however, can be constructed in the case at hand, due to the
absence of an $O(u)$ term in the square brackets for $\beta_{w}$ in
(\ref{betas2}). A regular solution can be constructed if we assume that
$y=O(\eps)$, but their difference $y-\eps$ is of order $\eps^{2}$. In other
words, the actual expansion parameters appear to be $\eps$ and
$\sqrt{(\eps-y)}$. It is reminiscent of the well known $\sqrt{\eps}$
expansion for the Ising ferromagnet with quenched disorder \cite{quench}.
There it was a consequence of an accidental degeneracy of the $\beta$
functions of the two involved couplings, which for their ratio (the analog
of our $w$) also implies vanishing of the first-order contribution.
A similar situation was also encountered in \cite{Satten} where
the mixing of a conserved order parameter by a Gaussian velocity ensemble
was studied.

To be definite, let us write $y=\eps+B \eps^{2}$ with some $B$. Now we have
\begin{eqnarray}
\beta_{u} &=& u [-\eps + au (1+w) +O(u^{2}) ],
\nonumber \\
\beta_{w} &=& w [-B\eps^{2} + bu^{2}(1+w)^{3}  +O(u^{3}) ]
\label{betasB}
\end{eqnarray}
and the solutions for $u_{*}$ and $w_{*}$ can be found as regular series
in $\eps$, while the dependence on $B$ should be taken into account
exactly in each order
of the $\eps$ expansion. Now we can identify two nontrivial fixed points,
which we denote as I and II.

\begin{figure}
\begin{center}
\includegraphics[width=10cm]{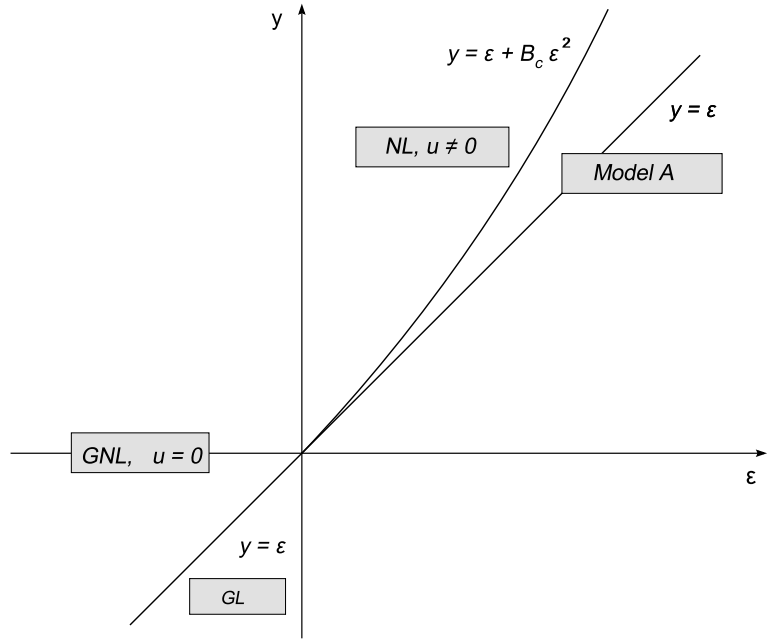}
\caption{\label{fig2} Regions of stability of the fixed points of
the extended model (\ref{model1}), (\ref{model}). The region
labelled GNL (Gaussian nonlocal) corresponds to the fixed point
with vanishing interaction ($d>4$) and purely nonlocal correlation
function of the noise; the label NL, $u\ne0$ refers to the
nonlocal regime II of the extended model; the first-order boundary
between the latter and the region of stability of the universality
class of model A (the local regime I of the extended model) is
$y=\varepsilon$ and the improved boundary is
$y=\varepsilon+B_{c}\varepsilon^{2}$. The boundary between the
universality class of model A and that of the linear model driven
by the random noise with local correlations only -- labelled GL
(Gaussian local) -- is the negative $y$ axis.}
\end{center}
\end{figure}

For the point I, we find $w_{*}=0$, $u_{*}=\eps/a$. This point clearly
corresponds to the model A of critical dynamics; $w_{*}$ vanishes
identically to all orders of the $\eps$ expansion (the model with
$w_{0}=w=0$ is local and therefore closed with respect to the
renormalization), while $u_{*}$ has nontrivial corrections of order
$\eps^{2}$ and higher. The $\Omega$ matrix at this point is:
\begin{eqnarray}
\beta_{uu} &=& au_{*}(1+w_{*}) = \eps,
\nonumber \\
\beta_{uw} &=& au^{2}_{*} = \eps^{2}/a, \nonumber \\
\beta_{wu} &=& 2u_{*}b \, w_{*}(1+w_{*}) =0,    \nonumber \\
\beta_{ww} &=& -B \eps^{2}+ bu^{2} = -B \eps^{2}+ b(\eps/a)^{2}.
\label{OmegaA}
\end{eqnarray}
It is triangular, so the fixed point is IR-attractive if the diagonal
elements $\beta_{uu}$ and  $\beta_{ww}$ are positive. This gives
$\eps>0$ (of course) and $B<b/a^{2}$. Thus we have established that
the region of IR-stability of the fixed point I includes not only the
sector $\eps>0$, $\eps>y$, but is slightly wider: it also involves a
narrow ``beak'' adjacent to the ray $\eps=y$ in the region $\eps<y$; see
figure \ref{fig2}, where regions of stability of the nontrivial fixed
points of the model (\ref{model1}), (\ref{model}) are depicted.

Furthermore, we can see that the fixed point with both $u_{*}\ne0$ and
$w_{*}\ne0$ also exists for the beta functions (\ref{betasB}); we denote
it as point II.
Indeed, $\beta_{u_{*}}=0$ and $u_{*}\ne0$ gives $u_{*}(1+w_{*})=\eps/a$,
substituting in $\beta_{w_{*}}=0$ and assuming $w_{*}\ne0$ gives
$$
0= -B \eps^{2} + b (1+w_{*}) [u_{*}(1+w_{*})]^{2} = \eps^{2} \left\{
-B + b(1+w_{*})/a^{2} \right\},
$$
so that at the fixed point
\begin{eqnarray}
w_{*}=-1+ Ba^{2}/b, \qquad u_{*}= \eps b/Ba^{3}
\label{FPB}
\end{eqnarray}
with corrections of order $O(\eps)$ and $O(\eps^{2})$ respectively.

The $\Omega$ matrix at this point is:
\begin{eqnarray}
\beta_{uu} &=& au_{*}(1+w_{*}) = \eps,
\nonumber \\
\beta_{uw} &=& au^{2}_{*} = \eps^{2}b^{2} /B^{2}a^{5}, \nonumber \\
\beta_{wu} &=& 2u_{*}b \, w_{*}(1+w_{*})^{3}
= 2bw_{*} (\eps/a)\, (a^{2}B/b)^{2} ,
\nonumber \\
\beta_{ww} &=& 3w_{*}bu^{2}_{*}(1+w_{*})^{2} = 3bw_{*} (\eps/a)^{2}.
\label{OmegaB}
\end{eqnarray}
Thus the $\Omega$ matrix has the form
\begin{equation}
\Omega = \eps \,
\left( \matrix { \alpha & {\cal A}\eps \cr
\beta & {\cal B}\eps \cr } \right)  .
\label{Omega}
\end{equation}
Although the elements in the right column are small in $\eps$ in
comparison to the left column, they are needed to find the eigenvalues
to leading order. It is important that the $O(\eps^{2})$ corrections
to the left elements are not needed (they give only corrections),
despite the fact that they would be of the same order as the right elements.

The eigenvalues are:
\begin{eqnarray}
\Omega_{1} = \alpha \eps +O(\eps^{2}), \qquad
\Omega_{2} = ({\cal B} -\beta {\cal A} /\alpha) \eps^{2} +O(\eps^{3}).
\label{lambdaB}
\end{eqnarray}
The point is IR-attractive if $\Omega_{1,2}>0$. Substituting
(\ref{OmegaB}), (\ref{FPB}) gives: $\eps>0$, $B>b/a^{2}$.

Thus the curve $y=\eps+B_{c}\eps^{2}$ with $B_{c}=b/a^{2}$ is the boundary
between the regions of IR-stability for the point I ($\eps>0$, $B<b/a^{2}$)
and the new nonlocal point II ($\eps>0$, $B>b/a^{2}$). There is neither gap
nor overlap (at least in this approximation).

The physics requires that the coordinates of any physical fixed point,
$u_{*}$ and $w_{*}$, be non-negative ($u$ and $uw$ are amplitudes in pair
correlators).
One can check that this condition is automatically satisfied in the region of
their IR stability. For example, $w_{*}>0$ in (\ref{FPB}) gives $B>b/a^{2}$,
which is also $\Omega_{2}>0$, and so on.

The resulting pattern of the regions of stability of the nontrivial fixed
points of the model (\ref{model1}), (\ref{model}), shown in figure
\ref{fig2}, is as follows:
the quadrant $\eps>0$, $y>0$ is divided into two parts by the parabola
$y=\eps+B_{c} \eps^{2}$ with $B_{c}=b/a^{2}$. The part below it is the region of
IR stability of the point I (universality class of the equilibrium model A);
it includes the whole sector $\eps>0$, $\eps>y$. The part above it is the
region of IR stability of the point II. It corresponds to a new
nonequilibrium universality class, where the nonlocal term in the random
force is important. It is worth noting that $B_{c}$ appears rather small:
$b=\ln (4/3)\approx 0.287683$ and $B_{c}= \ln(4/3)/9 \approx 0.032$.

\section{Scaling behaviour in the IR range} \label{sec:scaling}

Existence of IR-attractive fixed points implies scaling behaviour with
definite critical dimensions $\Delta_{F}$ of all quantities $F$
(fields and parameters):
\begin{eqnarray}
\Delta_{F} = d^{k}_{F}+ \Delta_{\omega} d^{\omega}_{F} + \gamma_{F}^{*},
\qquad  \Delta_{\omega}=2+\gamma_{\sigma}^{*},
\label{dim}
\end{eqnarray}
where $d^{k,\omega}_{F}$ are the canonical dimensions of $F$, given in
table~1, and $\gamma_{F}^{*}$ is the value of $\gamma_{F}$ at the fixed
point in question; see e.g. \cite{Book3,Red}. In the case at hand
$\gamma_{F}^{*}=\gamma_{F} (u_{*},w_{*})$.

In particular, for the pair correlator of the field $\varphi$ this gives:
\begin{eqnarray}
\langle \varphi (\x+\r, t+t') \varphi (\x, t') \rangle =
r^{-2\Delta_{\varphi}} \,
{\cal F} \left( \tau_{0}\,{r^{\Delta_{\tau}}}, \,
t/r^{\Delta_{\omega}} \right).
\label{scaling1}
\end{eqnarray}
This is the leading term of the asymptotic behaviour in the IR range,
determined by the inequalities $k \sim 1/r \ll \Lambda$, where $\Lambda$
is the UV momentum scale defined by the relation $u_{0} \sim \Lambda^{y}$.
${\cal F}$ is a universal scaling function of two arguments, which
are supposed to be of order unity; this completes the definition of the
IR range: $\tau_{0}\,{r^{\Delta_{\tau}}} \sim 1$,
$t/r^{\Delta_{\omega}} \sim 1$.
(In the free theory we would have $\tau_{0}r^{2}\sim1$, $t/r^{2}\sim1$).
The correct canonical dimensions in (\ref{scaling1}) are guaranteed by
the amplitudes built from IR irrelevant parameters $\sigma$ and $\Lambda$,
not shown explicitly.
One usually assumes that ${\cal F}$ has finite limits for $\tau_{0}=0$
(that is, exactly at the critical point) and/or $t=0$ (equal-time
correlator). Then from (\ref{scaling1}) one obtains
\begin{eqnarray}
\langle \varphi(\x+\r,t)\varphi(\x,t) \rangle = r^{-2\Delta_{\varphi}} .
\label{scaling2}
\end{eqnarray}

From the general expression (\ref{dim}), canonical dimensions
$d^{k,\omega}_{F}$ from the table \ref{table1} and the relations
(\ref{sol}) we obtain
\begin{eqnarray}
\Delta_{\varphi} = 1 - \eps/2 + (\gamma_{3} + \gamma_{2} - \gamma_{1})/2,
\quad \Delta_{\omega} = 2 + \gamma_{2} -\gamma_{3},
\nonumber \\
\Delta_{\varphi'} = 3 - \eps/2 + (\gamma_{1} + \gamma_{2} -\gamma_{3})/2,
\quad
\Delta_{\tau} = 2 -\gamma_{3} + \gamma_{5}.
\label{deltafi}
\end{eqnarray}

\begin{table}
\caption{\label{table1}Canonical dimensions of the fields and parameters;
$d=4-\varepsilon$.}
\begin{tabular}{@{}lllllllllllll}
\\
\hline
\mr
$F$ & $\varphi$ & $\varphi'$ &  $v_{i}$ & $v_{i}'$ & $\sigma_{0},\sigma$
& $\nu_0,\nu$ & $\!m,\mu,\Lambda\!$ & $\tau$ & $\lambda_{0}$ & $w_{0}$
& $g_{0}$ & $\lambda,w,g$ \\
\mr
$d_{F}^{k}$ & ${d\over 2}-1$ & ${d\over 2}-1$ & $-1$ & $\!d+1\!$ & 2 & $-2$ & 1 & 2 & $4-d$
& $d-4+y$ & $y'$ & 0 \\ $d_{F}^{\omega}$  & 0 & 1 & 1 & $-1$ &
$-1$ & 1 & 0 & 0 & 0 &  0  &  0  &  0 \\
$d_{F}$ & ${d\over 2}-1$ & ${d\over 2}+1$ & 1 & $\!d-1\!$ & 0  &  0  & 1 & 2 &
$4-d$ & $d-4+y$ & $y'$ & 0 \\
\mr\hline
\end{tabular}
\end{table}

In the leading approximation the anomalous dimensions have the forms
\begin{eqnarray}
\gamma_{1} = b\, u^{2}(1+w)^{3}, \quad
\gamma_{2} = b\, u^{2}(1+w)^{2}, \quad
\gamma_{3} = u^{2}(1+w)^{2}  /6, \nonumber \\
\gamma_{4} = -3 u(1+w), \quad
\gamma_{5} = - u(1+w)
\label{NL2}
\end{eqnarray}
with $b=\ln (4/3)$. For $w=0$ these expressions coincide with the results
known for the model A; see e.g. \cite{Book3}.

Consider first the fixed point I with $u_{*}=\eps/a+O(\eps^{2})$ and
$w_{*}=0$.
The local model ($w=0$) does not generate nonlocal counterterms and is
``closed with respect to renormalization.'' As a result, $w_{*}$ vanishes
identically, the dimensions (\ref{deltafi}) are independent of $y$ and
coincide with their analogs of the model A to all orders of the $\eps$
expansion. For the model A, the fluctuation-dissipation theorem gives the
exact relation $\gamma_{1}^{*} = \gamma_{2}^{*}$, so that these quantities
disappear from the expression for $\Delta_{\varphi}$ in
(\ref{deltafi}), while the dimensions $\gamma_{\varphi,\tau,u}^{*}$
(and hence $\Delta_{\varphi,\tau}$) coincide with their counterparts in
the static $\lambda\varphi^{4}$ model; see e.g. \cite{Book3,Zinn}.

The standard notation for the equilibrium case is
\begin{eqnarray}
\Delta_{\varphi}=d/2-1+\eta/2, \quad \Delta_{\tau}=1/\nu, \quad
\Delta_{\omega}=z,
\label{static}
\end{eqnarray}
while for $\varphi'$ the above identities give $\Delta_{\varphi'}=
\Delta_{\varphi}+z=d/2-1+z+\eta/2$. The static exponents $\eta$ and $\nu$
are well known from $4-\eps$, $2+\eps$ and $1/N$-expansions, real-space RG
(all augmented by various summations), high-temperature expansions for the
Ising model (considered most reliable), Monte-Carlo simulations. The values
recommended by \cite{Book3,Zinn}  are $\eta=0.0375 \pm 0.0025$ and
$\nu= 0.6310\pm0.0015$ (Borel summation of 5-order results).
For $z$ only two terms of the $4-\eps$ expansion are known:
$z= 2+0.726 (1-0.1885 \eps)\eta$ \cite{AV}; there are also leading-order
results in $2+\eps$ and $1/N$-expansions
(see the references in \cite{HH,FM,Book3}).

To avoid possible misunderstandings, it is worth noting that the scaling
behaviour of the model A and the extended model (\ref{MSRR}) near its fixed
point I coincide only to the leading orders given by the expressions
(\ref{scaling1}), (\ref{scaling2}). The corrections to those expressions
are different. In particular, the leading correction from the UV range has
the form $\left[1+ c_{1} (k/\Lambda)^{\Omega_{1}} + c_{2}(k/\Lambda)^
{\Omega_{2}} \right]$, where $\Omega_{1,2}$ are the eigenvalues of matrix
$\Omega$ for the fixed point I given in (\ref{OmegaA}). One of them
involves the parameter $y$ which is absent in the pure model A.

Let us turn to the nonlocal regime of the extended model (\ref{MSRR}),
described by the fixed point II. Substituting (\ref{NL2}) and (\ref{FPB})
into (\ref{deltafi}) gives
\begin{eqnarray}
\Delta_{\varphi} &=& 1 -\eps/2 + (\eps^{2}/2) (1/54+b/9  - B)+O(\eps^{3}),
\nonumber \\
\Delta_{\varphi'} &=& 3 -\eps/2 + (\eps^{2}/2) (-1/54+b/9  - B)+O(\eps^{3}),
\nonumber \\
\Delta_{\omega} &=& 2+ (\eps/3)^{2} (b-1/6)+O(\eps^{3}),
\nonumber \\
\Delta_{\tau}  &=& 2-\eps/3+O(\eps^{2}).
\label{deltafi2}
\end{eqnarray}
We recall that $B$ comes from the relation $y=\eps+B\eps^{2}$ and the
nonlocal fixed point is IR-attractive when $B>b/9\approx 0.032$, see
(\ref{lambdaB}). Note that the dimension $\Delta_{\omega}=2+ 0.0013
\eps^{2}+O(\eps^{3})$ appears surprisingly close to its canonical value
$\Delta_{\omega}=2$.

Consider the real case $d=3$, then $y=1+B$ and the noise correlator is
$1/r^{3-B}$. The exponent $\eta$, defined by the same ``equilibrium''
relation
$\Delta_{\varphi}=d/2-1+\eta/2$, takes on the form $\eta=(1/54+b/9-B)\approx
0.05-B$. So $\eta$ can be made negative for reasonable $B$: in particular,
for $B=1$ and noise correlator $1/r^{2}$ we have $\eta\approx -1$. In this
respect, the nonequilibrium steady-state scaling differs from the
equilibrium case, described by a local Landau--Ginzburg action: for the
latter, the exact inequality $\eta>0$ can be derived from the unitarity
of the corresponding pseudo-Euclidean quantum field theory \cite{Pol}.
Bearing in mind possible cosmological application of the model \cite{Inflat}, it is
tempting to note that $\eta=-1$ corresponds to the Zeldovich spectrum
\cite{Zel}.

\section{Inclusion of the velocity field} \label{sec:velocity}

Let us turn to the full stochastic problem (\ref{1.1}), (\ref{model1}),
(\ref{model}). The field theoretic action functional then becomes
\begin{equation}
S(\Phi)= S_{v}(v',v) + \varphi' D_{\varphi} \varphi' /2 +
\varphi' \left\{- \sigma_{0}\nabla_{t} \varphi + \partial^{2} \varphi
- \tau_{0} \varphi - \lambda_{0} \varphi^{3} /6 \right\},
\label{actionF}
\end{equation}
where
\begin{equation}
S_{v}(v',v)= v 'D_{v} v'/2+ v' \left\{-\nabla_t +\nu _0\partial^{2} \right\}v
\label{actionV}
\end{equation}
is the action functional for the stochastic problem (\ref{1.1}),
$D_{\varphi}$ and $D_{v}$ are the correlation functions of the random forces
$f$ and $f_{i}$, respectively, $\nabla_{t}=\partial_{t}+v_{i}\partial_{i}$,
and all the required integrations over $x=\{t,{\bf x}\}$ and summations over
the vector indices are understood. The new full set of fields
$\Phi=\{ \varphi, \varphi',  v_{i}, v'_{i} \}$ involves the auxiliary vector
field $v_{i}'$. It is also transverse, $\partial_{i}v_{i}'=0$,
which allows one to omit the pressure term on the right-hand side of relation
(\ref{actionV}). Correlation function $D_{\varphi}$ is given by
(\ref{model}), while $D_{v}$ will be taken in the form
\begin{eqnarray}
\langle f_{i}(x)f_{j}(x') \rangle &=& g_{0}\nu_{0}^{3}\, \delta(t-t')
k^{4-d-y'} P_{ij}({\bf k}),
\label{model2}
\end{eqnarray}
where $P_{ij}({\bf k}) =\delta _{ij}- k_i k_j / k^2$ is the transverse
projector, $y'$ a new arbitrary parameter analogous to $y$ from
(\ref{model}), $g_{0} \sim \Lambda^{y'}$ is a new positive coupling constant,
and the factor $\nu_{0}^{3}$ is explicitly isolated for convenience.
The IR cutoff at $k=m$ is also implied.
Canonical dimensions of all the new parameters and their future renormalized
counterparts are given in table~1.

The stochastic Navier--Stokes equation (\ref{1.1}) with a power-law noise
spectrum was introduced a long ago \cite{FNS,DeDom,Frisch} and is by now
very well studied, at least for small values of $y'$. The two-loop results
have been derived recently in \cite{Komp}. Detailed exposition of the RG
approach can be found in \cite{Book3,Red}; below we confine ourselves to
only the necessary information.

The model (\ref{actionV}) is logarithmic (the coupling constant $g_0$ is
dimensionless) at $y'=0$, and the UV divergences have the form of the poles
in $y'$ in the correlation functions of the fields $v$, $v'$.
Dimensional analysis, augmented by some additional considerations
(Galilean symmetry and structure of the vertex), shows that for
all $d>2$, the superficial UV divergences, whose removal requires
counterterms, are present only in the 1-irreducible function
$\langle v'v \rangle$, and the corresponding counterterm reduces
to the form $v'\partial^{2}v$. Owing to the form of the vertex
(the derivative can always be moved onto $v'$ using integration by parts),
the divergence in the function $\langle v' v'\rangle$ (allowed
by dimension for $d<4$) is in fact absent for all $d>2$. So the local
counterterm $v'v'$, analogous to $\varphi'\varphi'$ in (\ref{MSRR}),
is not needed here. For this reason, we did not include the constant
contribution to (\ref{model2}), in contrast to its scalar counterpart
(\ref{model}). Then for the complete elimination of the UV divergences it
is sufficient to perform the multiplicative renormalization of the parameters
$\nu_0$ and $g_{0}$ with the only independent renormalization constant
$Z_{\nu}$:
\begin{equation}
\nu_0=\nu Z_{\nu}, \qquad g_{0}=g\mu^{y'}Z_{g},
\qquad Z_{g}=Z_{\nu}^{-3} .
\label{18}
\end{equation}
Here $\mu$ is the reference mass in the MS scheme, $g$ and $\nu$ are
renormalized analogs of the bare parameters $g_{0}$ and $\nu_0$, and
$Z=Z(g,y',d)$ are the renormalization constants. In contrast to the model
(\ref{MSRR}), no renormalization of the fields is needed, $Z_{v.v'}=1$.
The relation between the $Z$'s in (\ref{18}) results from the absence of
renormalization of the noise term in (\ref{actionV}). Now the standard RG
equations are readily derived, the corresponding function $\beta_{g}=\Dm g$
in the one-loop approximation is
\begin{equation}
\beta_{g}(g)= g(-y'+3\gamma_{\nu})=
g\left\{-y'+ \frac{3(d-1) S_{d}\,g}{ 4(d+2) (2\pi)^d}
\right\}+O(g^3)
\label{26}
\end{equation}
where $S_{d}=2\pi^{d/2}/\Gamma(d/2)$ is the area of the unit sphere in
$d$ dimensions. From (\ref{26}) we immediately conclude that for $y'>0$,
the model has an IR attractive nontrivial fixed point $g_{*}>0$
($\beta_{g}(g_{*})=0$, $\beta_{g}'(g_{*})>0$), while for $y'<0$ the IR
attractive fixed point is Gaussian, $g_{*}=0$ (that is, the nonlinearity
in (\ref{1.1}) is IR irrelevant). From the first equality in (\ref{26}),
which follows from the relation between the $Z$'s in (\ref{18}), the value
of $\gamma_{\nu}$ at the nontrivial fixed point is found exactly:
$\gamma_{\nu}^{*}=y'/3$ (no corrections of order $(y')^{2}$ and higher).
As a result, critical dimensions of the frequency and the fields are also
found exactly:
\begin{equation}
\Delta_{\omega}=2-y'/3, \quad \Delta_{v}=1-y'/3, \quad \Delta_{v'}=d-1+y'/3.
\label{7.40}
\end{equation}

These results remain intact in the full model (\ref{actionF}), because the
inclusion of the scalar fields does not affect the velocity (the field
$\varphi$ is ``passive''). We thus conclude that for $y'<0$, where the IR
behaviour of the velocity becomes Gaussian, the scaling regimes of the full
problem (\ref{actionF}) are described by the fixed points I and II from
section~\ref{sec:Fixed} depending on the relation between $y$ and $\eps=4-d$.

For $y'>0$ the contributions of the velocity field become IR relevant,
and the RG analysis of the full problem (\ref{actionF}), (\ref{actionV}) is
needed. For $d>4$, where no local counterterm $\varphi'\varphi'$ is required,
the model in question is formally equivalent to the stochastic model of the
turbulent advection of a chemically active scalar field, studied earlier
in~\cite{IJMP}. We will be interested in the case $d<4$, where the
local counterterm should be included from the very beginning to ensure
multiplicative renormalizability. Then the UV divergences manifest
themselves as poles in the full set of regulators $y$, $y'$ and $\eps=4-d$.
Dimensional analysis and symmetry considerations show that, for all $d>2$,
the full model is multiplicatively
renormalizable, and the corresponding renormalized action has the form
\begin{eqnarray}
S_{R}(\Phi) &=& S_{vR}(v',v) + \varphi' \left\{ w\mu^{y-\eps} k^{\eps-y}
+ Z_{1} \right\} \varphi' /2
\nonumber \\
&+& \varphi' \left\{- Z_{2}\sigma \nabla_{t} \varphi +
Z_{3}\partial^{2} \varphi - \tau Z_{5} \varphi - Z_{4}\lambda \mu^{\eps}
\varphi^{3} /6 \right\},
\label{actionFR}
\end{eqnarray}
where $S_{vR}(v',v)$ is the renormalized analog of the action (\ref{actionV})
expressed in renormalized variables using relation (\ref{18}). The renormalization
constants $Z_{1}$--$Z_{5}$ now contain the poles in the full set of
regulators $y$, $y'$, $\eps=4-d$ and, in comparison to (\ref{MSRR}), depend
on two additional dimensionless couplings $g$ and $s=\sigma\nu$. One can
easily see that the relations (\ref{reno}), (\ref{Z}) for the new constants
$Z_{F}$ and (\ref{Z1}), (\ref{sol}) for the corresponding anomalous
dimensions $\gamma_{F}$ remain valid in the extended renormalized model
(\ref{actionFR}). The RG equation takes on the form
\begin{equation}
\left\{ {\cal D}_{RG} + N_{\varphi}\gamma_{\varphi} + N_{\varphi'}
\gamma_{\varphi'}
+ N_{v}\gamma_{v} + N_{v'}\gamma_{v'} \right\} \,G^{R}(e,\mu,\dots) = 0,
\label{RG1F}
\end{equation}
where ${\cal D}_{RG}$ is the operation $\widetilde{\cal D}_{\mu}$
expressed in the renormalized variables:
\begin{equation}
{\cal D}_{RG}\equiv {\cal D}_{\mu} + \beta_{g}\partial_{g} +
\beta_{u}\partial_{u} + \beta_{w}\partial_{w} + \beta_{s}\partial_{s}
- \gamma_{\nu}{\cal D}_{\nu} - \gamma_{\tau}{\cal D}_{\tau},
\label{RG2F}
\end{equation}
with $\beta_{g}$ from (\ref{26}), $\beta_{u,w}$ from (\ref{Betas}),
and the new function $\beta_{s}$ is
\begin{equation}
\beta_{s} = -s\gamma_{s}= -s[\gamma_{\nu}+\gamma_{\sigma}]=
-s[\gamma_{\nu}+\gamma_{2}-\gamma_{3}].
\label{betav}
\end{equation}
Furthermore, it is easily checked that in the leading order, the anomalous
dimensions $\gamma_{1,2,4,5}$ remain the same as in the model without
velocity and can be taken from (\ref{NL2}). On the contrary, $\gamma_{3}$
acquires an additional $O(g)$ term, in comparison to which the $O(u^{2})$
term in (\ref{NL2}) is only a correction and should be neglected. So the
leading-order expression for $\gamma_{3}$ becomes
\begin{equation}
\gamma_{3}= \frac{3gs^{2}}{64\pi^2 (s+1)} +O(g^{2},u^{2}).
\label{NL2F}
\end{equation}

As usual, scaling regimes of the full model are associated with the IR
attractive fixed points, whose coordinates are found from the equations
$\beta_{x}=0$ with $x=\{g,s,u,w\}$. Due to passivity of the field
$\varphi$, the function $\beta_{g}$ is independent of $x=\{s,u,w\}$ and
the corresponding elements $\partial\beta_{g}/\partial x$ of the matrix
$\Omega$ vanish. Thus $\Omega$ is triangular, its elements
$\partial\beta_{x}/\partial g$ with $x=\{s,u,w\}$ do not affect the
eigenvalues, and we can set $g=g_{*}$ in the functions $\beta_{x}$
with $x \ne g$ from the very beginning. Now we are treating $\eps=4-d$
as one of the small expansion parameters, and in the leading-order
approximation we should set $d=4$ in (\ref{26}) and $g_{*}$.
In the leading approximation, the function $\beta_{s}$ is independent
of $u$ and $w$, and the elements $\partial\beta_{s}/\partial x$ with
$x=\{u,w\}$ also vanish. The element $\partial\beta_{s}/\partial s$
coincides with one of the eigenvalues of $\Omega$. It is easy to see that,
for $y'>0$, this eigenvalue $\partial\beta_{s}/\partial s$ cannot be
positive for $s_{*}=0$, so that the condition $\beta_{s}=0$ implies
$\gamma_{\nu}+\gamma_{2}-\gamma_{3}=0$, which in our approximation gives
$3s_{*}^{2}-s_{*}-1=0$. Physical considerations require $s_{*}>0$, which
finally gives $s_{*}=(1+\sqrt{13})/6 \approx 0.7676$. Since the elements
$\partial\beta_{x}/\partial s$ with $x=\{u,w\}$ are again irrelevant in the
analysis of the remaining eigenvalues, we can set $s=s_{*}$ in $\beta_{u,w}$.
We are left with the system of two equations, $\beta_{u,w}=0$, where
\begin{eqnarray}
\beta_{u} &=& u [-\bar\eps + au (1+w) +O(g^{2},u^{2}) ],
\nonumber \\
\beta_{w} &=& w [-\bar y+\bar\eps + bu^{2}(1+w)^{3}  +O(g^{3},u^{3}) ].
\label{betasF}
\end{eqnarray}
Here we have denoted
\begin{equation}
\bar\eps= \eps-2y'/3, \qquad \bar y = y -2y'/3.
\label{bars}
\end{equation}
Up to the notation, this system coincides with that describing the model
without velocity, equation (\ref{betas2}), and its fixed points and their
regions of stability can immediately be inferred from the results of
section~\ref{sec:Fixed}. In order to have a regular expansion we again set
$\bar y=\bar\eps+\bar B \bar\eps^{2}$, while $\bar\eps$ and $y'$ are of the
same order. (From the relation $\bar\eps-\bar y=\eps -y$ one obtains
$B\eps^{2}=\bar B \bar \eps^{2}$. We could also write $y'=B' \bar\eps$,
but this is not necessary.)
Now we can identify two fixed points, which we denote III and IV,
corresponding to completely new universality classes.

The point III corresponds to a purely local model: $w_{*}=0$ to all orders
of the perturbative expansion, and $u_{*}=\bar\eps/a$ with corrections
$O(\bar\eps^{2})$. It is IR attractive for $\bar\eps>0$ and $\bar B<b/a^{2}$
(we recall that $a=3$ and $b=\ln (4/3)$). For the point IV, one obtains
$w_{*}= -1+ \bar B a^{2}/b$ and $u_{*}= \bar\eps b/\bar Ba^{3}$, with
corrections of order $O(\bar\eps)$ and $O(\bar\eps^{2})$ respectively.
It is IR attractive for $\bar\eps>0$, $\bar B>b/a^{2}$. The curve
$\bar y=\bar\eps+\bar B_{c}\bar\eps^{2}$ with $\bar B_{c}=b/a^{2}
\approx 0.032$ is the boundary between the regions of IR-stability for
these two points.

For both these regimes, the relations $\beta_{v}=0$ and $v_{*}\ne0$ imply
\begin{equation}
\gamma^{*}_{\sigma}= \gamma^{*}_{2}-\gamma^{*}_{3}= -\gamma^{*}_{\nu}=-y'/3,
\label{REL}
\end{equation}
and the critical dimension of frequency is found exactly:
\begin{equation}
\Delta_{\omega}= 2+\gamma^{*}_{\sigma} = 2-\gamma^{*}_{\nu} =2-y'/3,
\label{Fri}
\end{equation}
in agreement with (\ref{7.40}). Combining Eqs. (\ref{sol})
and (\ref{REL}) gives
\begin{eqnarray}
2 \gamma^{*}_{\varphi} = \gamma^{*}_{3}+\gamma^{*}_{2}-\gamma^{*}_{1}=
\gamma^{*}_{\nu}+2\gamma^{*}_{2}-\gamma^{*}_{1},\nonumber
\\
2 \gamma^{*}_{\varphi'} = \gamma^{*}_{3}+\gamma^{*}_{1}-\gamma^{*}_{2}=
\gamma^{*}_{\nu}+ \gamma^{*}_{1},
\label{Xer}
\end{eqnarray}
which, along with (\ref{Fri}), allows one to determine the dimensions
$\Delta_{\varphi} = (d/2-1)+\gamma^{*}_{\varphi}$ and
$\Delta_{\varphi'} = (d/2-1)+ \Delta_{\omega}+ \gamma^{*}_{\varphi'}$
in the second order of the $\eps$ expansion without practical calculation
of the two-loop corrections to $\gamma_{3}$ and $v_{*}$:
\begin{equation}
\Delta_{\varphi} = 1-\eps/2+y'/6 +b\bar\eps^{2}/18, \quad
\Delta_{\varphi'} = 3-\eps/2 -y'/6+b\bar\eps^{2}/18
\label{Dtri}
\end{equation}
for the fixed point III and
\begin{equation}
\Delta_{\varphi} = 1-\eps/2+y'/6 +b\bar\eps^{2}/9-\bar B \bar\eps^{2}/2,
\quad \Delta_{\varphi'} = 3-\eps/2 -y'/6+\bar B \bar\eps^{2}/2
\label{DtriX}
\end{equation}
for the fixed point IV, with cubic-in-$\eps$ corrections. (We recall
that in counting the orders we imply $\eps\sim\bar\eps\sim y'$,
$\eps-y\sim\eps^{2}$). In fact, the expression for $\Delta_{\varphi'}$
in (\ref{DtriX}) holds to all orders (no corrections of order $\eps^{3}$
and higher), because $\gamma_{1}^{*}=y-\eps =B \eps^{2}$ exactly, as a
consequence of the relations $\beta_{w}=0$ and $w_{*}\ne 0$. Finally, for
the both points III and IV one obtains $\Delta_{\tau}=2+ \eps/3 +y'/9$, with
quadratic corrections.

For $\bar\eps<0$, the self-interaction of the scalar field becomes
irrelevant ($u_{*}=0$), and we obtain two more fixed points, which
correspond to the scalar field subject to a linear diffusion-advection
equation, with the velocity ensemble given by the action $S_{v}(v',v)$
from (\ref{actionV}). The regions $\bar\eps>\bar y$ and $\bar\eps<\bar y$
correspond to the local ($w_{*}=0$) and nonlocal correlator of the
scalar noise, respectively.
For the purely nonlocal case, the amplitude
$w$ remains finite at the fixed point and can be eliminated from the
action by appropriate rescaling of the fields $\varphi$ and $\varphi'$.

The linear passive scalar case is very well understood within the RG
framework (see e.g. chapter~2 in the book \cite{Red} and references
therein). The only superficially divergent 1-irreducible correlation
function is $\langle \varphi' \varphi \rangle$ with the counterterm
$\varphi' \partial^{2} \varphi$, and the corresponding renormalization
constant (in our notation identified with $Z_{\varphi}=Z_{\sigma}^{-1}$)
is independent of the form of the noise correlator (the latter only
determines the canonical dimensions). The absence of renormalization of
the noise term results in the exact relation $Z_{\varphi'}=1$ (it is implied
that the amplitude is scaled out from the action). Along with the relations
(\ref{REL}) and (\ref{Fri}), which remain valid for the passive linear case,
this gives exact results for the dimensions:
\begin{equation}
\Delta_{\varphi} = d/2-1+y'/3, \quad \Delta_{\varphi'} = d/2+1-y'/3, \quad
\Delta_{\tau} = 2,
\label{LiPa}
\end{equation}
which in the standard ``equilibrium'' notation corresponds to
$\eta=y'/3$ and $\nu=1/2$, different from their counterparts for the
standard model A.

\section{Conclusion} \label{sec:Conc}

We have studied stochastic model that describes dynamics of a nonconserved
scalar field (order parameter) near its critical point, subject to random
external stirring and mixing, in $d$ spatial dimensions. The stirring was
modelled by an additive random Gaussian noise with the pair correlation
function $\propto\delta(t-t') k^{4-d-y}$. The mixing was modelled by
the convection term with a divergence-free (due to the incompressibility
condition) velocity field, governed by the stochastic Navier--Stokes
equation with a random Gaussian force with pair correlation function
$\propto\delta(t-t') k^{4-d-y'}$. Possible scaling regimes of the model
are associated with nontrivial IR attractive fixed points of the
corresponding RG equations. Their coordinates, regions of stability, and
the corresponding critical dimensions can be calculated within a systematic
expansion in $y$, $y'$ and $\eps=4-d$ (or only $y$ and $\eps=4-d$ for the
model without velocity) with the additional assumption that
$y-\eps =O(\eps^{2})$. Depending on the relations between those
parameters, the model reveals several types of scaling regimes.
Some of them are well known: model A of equilibrium critical dynamics and
linear passive scalar field advected by a random turbulent flow, but
there are three new nonequilibrium universality classes, associated
with new nontrivial fixed points. In this sense, the critical behaviour of
the model appears richer and less universal than that of the equilibrium
critical dynamics.

The critical exponents (dimensions) for the new universality classes
are derived in the second order of the expansion in $y$, $y'$ and $\eps$
(two-loop approximation).

It remains to note that the large-scale mixing ($y=y'=4$) in three
dimensions ($\eps=1$) belongs to the universality class of the
linear passive scalar with the nonlocal noise correlator and
therefore corresponds to the dimensions (\ref{LiPa}). Of course,
the results of our perturbative RG analysis are absolutely
reliable and internally consistent only for small values of the
expansion parameters $\eps$, $y$ and $y'$, while the possibility
of their naive extrapolation to finite (and not small) real values
is far from obvious. On the other hand, the observation that the
$\varphi^{4}$-interaction becomes irrelevant for the large-scale
forcing is reminiscent of the results derived in
\cite{Akira,Beysens}. There, it was argued that a non-random shear
flow strongly suppresses critical fluctuations, and the behaviour
of the system becomes close to mean field in the strong shear
limit; see also discussion in \cite{Chan}.

Our analysis can be directly generalized to the cases of a $N$-component
order parameter, presence of anisotropy, compressibility etc. The
generalizations are straightforward but rather cumbersome (for the
stochastic Navier--Stokes equation, see e.g. chapter~3 in the book \cite{Red}
and references therein). On the contrary, the case of a conserved order
parameter appears rather different from both the conceptual and technical
viewpoints (namely, it involves two different dispersion laws:
$\omega \sim k^{2}$ for the velocity and $\omega \sim k^{4}$ for the
scalar). These issues will be addressed elsewhere.

\section*{Acknowledgments}

The authors thank L Ts  Adzhemyan, Massimo Cencini, Paolo Muratore
Ginanneschi, Filippo Vernizzi, Angelo Vulpiani and A N Vasil'ev
for discussions. N V A  was supported in part by the RFFI grant no
05-02-17\,524 and the RNP grant no~2.1.1.1112. M~H was supported
in part by the VEGA grant 6193 of Slovak Academy of Sciences, by
Science and Technology Assistance Agency under contract No
APVT-51-027904. N~V~A and M~H thank the Department of Physical
Sciences in the University of Helsinki and the N~N Bogoliubov
Laboratory of Theoretical Physics in the Joint Institute for
Nuclear Research (Dubna) for their kind hospitality. N~V~A  thanks
the Department of Mathematics in the University of Helsinki for
their kind hospitality during his visits, financed by the project
``Extended Dynamical Systems.''

\appendix
\section{Calculation of the renormalization constants} \label{sec:Zs}

Consider as an example the calculation of the constant $Z_{1}$ and the
anomalous dimension $\gamma_{1}$ for the model (\ref{model})
without the velocity field in detail. The leading contribution here is
given by a two-loop Feynman graph, so this example is representative:
calculation of the other renormalization constants (including the velocity
field) can be performed in a similar way (for two-loop contributions) or
is much easier (for one-loop graphs).

The 1-irreducible function $\Gamma\equiv \langle \varphi' \varphi' \rangle $
in the renormalized critical ($\tau=0$) theory to order $O(\lambda^{2})$ has
the form
\begin{eqnarray}
\Gamma = 2\sigma\,  \left\{ w (k/\mu)^{y-\eps} + Z_{1} \right\}
+ \frac{\lambda^{2}\mu^{2\eps}}{6} \times {\rm Diagram},
\label{Gamma}
\end{eqnarray}
where the first term is the noise correlator written in renormalized
variables, $Z_{1}$ should be taken to order $O(\lambda^2)$, 1/6 is the
symmetry coefficient, $\lambda^{2}\mu^{2\eps}$ comes from the vertex factors,
dashed external ends correspond to fields $\varphi'$ and the three
identical lines correspond to bare propagators
\begin{eqnarray}
\langle \varphi \varphi \rangle_{0} = \frac{ 2\sigma\,
\left\{ w\, (k/\mu)^{y-\eps} + 1 \right\} }
{ \omega^{2} \sigma^{2}+ k^{4}}
\label{lines}
\end{eqnarray}
in frequency-momentum representation and
\begin{eqnarray}
\langle \varphi \varphi \rangle_{0} =
\left\{ w\, (k/\mu)^{y-\eps} + 1 \right\}  \,
\frac{1}{k^{2}} \, \exp{ (-k^{2} |t-t'|/\sigma) }
\label{lines2}
\end{eqnarray}
in time-momentum representation. The second bare propagator
\begin{eqnarray}
\langle \varphi \varphi' \rangle_{0} = (-{\rm i}\sigma\omega+k^{2})^{-1}
\to \theta(t-t') \exp{ (-k^{2}(t-t')/\sigma) }
\label{lines3}
\end{eqnarray}
existing in the diagrammatic technique for the model (\ref{MSRR}) does not
appear in this diagram. Within our accuracy, all $Z$'s occurring in the
diagram (for example, $Z_{4}^{2}$ coming from renormalized vertices) have
been replaced with unities.

The constant $Z_{1}$ will be found from the condition that it cancels
the poles in $\eps$ and $y$ which are present in the diagram, so the full
expression (\ref{Gamma}) is regular in $\eps$ and $y$ and finite for
$\eps=0$,  $y=0$. This requirement determines $Z_{1}$ up to a regular part.
We will use the minimal subtraction (MS) scheme where $Z_{1}=1+$ only poles
in $\eps$, $y$ and their linear combinations. In the model with two
regulators like $\eps$, $y$ there are subtleties in defining the MS
scheme in higher orders (for example, is $\eps/y$ a pole or a finite
thing). We will work only in the leading order and can neglect these
subtleties.

It is sufficient to calculate the diagram with the external frequency and
momentum equal to zero. In the theory above the critical temperature
($\tau>0$) the IR regularization is provided by the replacement
$k^{2}\to k^{2}+\tau$ in the denominator of (\ref{lines2})). This becomes
impossible for the case $\tau\le0$, which we are mostly interested in here.
The adequate language is then provided by the Legendre transform (effective
action) and use of the loop expansion or the $1/N$ expansion instead of the
primitive perturbation theory. This is not convenient, however, for the
practical calculation of the renormalization constants. Fortunately, in
the MS scheme the counterterms are polynomial in IR regulators, and the
results obtained for them in the region $\tau>0$ can be directly used for
$\tau\le0$; see e.g. the discussion section~3.36 in \cite{Book3}. Furthermore,
the constants $Z$ are independent on the specific choice of the IR
regularization. From the calculational viewpoints, it is more convenient to
set $\tau=0$ in the action (and in the propagator (\ref{lines2})) and cut off
the momentum integrals at $k=m$ (by dimension, $\tau \sim m^{2}$). Integrals
over frequencies (or times) are elementary, and one obtains:
\begin{eqnarray}
{\rm Diagram}= {2\sigma}\!\! \int\limits_{k>m}\!\!\!\!
\frac{d\k}{(2\pi)^{d}} \,
\int\limits_{q>m} \!\!\!\!\frac{d\q}{(2\pi)^{d}} \, \frac {D(k)D(q)D(|\k+\q|)}
{k^{2}q^{2}(\k+\q)^{2} [k^{2}+q^{2}+(\k+\q)^{2}] },
\label{D1}
\end{eqnarray}
where we have denoted
\begin{eqnarray}
D(k) \equiv [w\, (k/\mu)^{y-\eps} + 1].
\label{DD}
\end{eqnarray}

The integrand in (\ref{D1}) depends only on three independent variables:
the moduli $k=\k$, $q=\q$ and the angle $\vartheta$ between the directions
$\k$ and $\q$, so that $\k\q=kq\cos\vartheta$. Thus the expression (\ref{D1})
can be written as a linear combinations of the integrals of the form
\begin{eqnarray}
\int\limits_{k>m}\!\!\!\! \frac{d\k}{(2\pi)^{d}}\,\int\limits_{q>m} \!\!\!\! \frac{d\q}{(2\pi)^{d}}
\dots = \nonumber \\ {}  \nonumber \\
\frac{S_{d}^{2}}{(2\pi)^{2d}} \,
\int\limits_{m}^{\infty}\!\! \frac {dk}{k^{1+\eps}}
\int\limits_{m}^{\infty}\!\! \frac {dq}{q^{1+\eps}} \langle\langle
\frac {k^{2}q^{2}k^{\alpha_{1}}q^{\alpha_{2}}|\k+\q|^{\alpha_{3}}}
{(\k+\q)^{2} [k^{2}+q^{2}+(\k+\q)^{2}]} \rangle\rangle
\label{D2}
\end{eqnarray}
where each $\alpha_{i}$ is either 1 or $\eps-y$, the brackets mean the
averaging over the angle $\vartheta$ normalized as
$\langle\langle 1 \rangle\rangle=1$ and
$S_{d}=2\pi^{d/2}/\Gamma(d/2)$ is the area of the unit sphere in the
$d$-dimensional space.

From dimensional considerations it is obvious that for the integrals $I(m)$
in (\ref{D2}) we have
\begin{eqnarray}
I(m) = m^{-2\eps+\alpha_{123}} I(m=1), \qquad
\alpha_{123}= \sum_{i} \alpha_{i}.
\label{D3}
\end{eqnarray}
Thus
\begin{eqnarray}
{\cal D}_{m}I(m) = (-2\eps+\alpha_{123}) I(m), \qquad {\cal D}_{m}\equiv
m \partial_{m}
\label{D4}
\end{eqnarray}
and
\begin{eqnarray}
I(m) = - m^{-2\eps+\alpha_{123}} \frac{1} {(2\eps-\alpha_{123})}
{\cal D}_{m}I(m)|_{m=1}.
\label{D5}
\end{eqnarray}
Here the pole is isolated explicitly. The expression
${\cal D}_{m}I(m)|_{m=1}$ is finite at $\eps=y=0$, and we can set
$\eps=y=0$ in it. Then all these integrals become equal (all $\alpha_{i}$
also become 0).

The factors $m^{-2\eps+\alpha_{123}}$ will form dimensionless ratios like
$(m/\mu)^{O(\eps)}$ or $(m/\mu)^{O(y)}$ with the $\mu$-dependent factors
in expressions (\ref{Gamma}) and (\ref{DD}). Since we are interested only
in the pole parts we will replace such ratios by unities.
Thus we have to calculate
\begin{eqnarray}
{\cal R} = - {\cal D}_{m}I(m)|_{m=1,\eps=y=0}.
\label{D6}
\end{eqnarray}
Since $m$ appears in $I(m)$ only in the lower limits of integration,
the differentiation gives
\begin{eqnarray}
{\cal R} = \int\limits_{1}^{\infty} dq \, \langle\langle\,
\frac{q}{(1+q^{2}+2q\cos\vartheta)(1+q^{2}+q\cos\vartheta)}
\,\rangle\rangle.
\label{D10}
\end{eqnarray}
Here we have rescaled the variable $q/m\to q$, so that new $q$ is
dimensionless and $m$ has disappeared. The factor 2 (there are two
equal contributions
since the integrand was symmetrical in $q$ and $k$) cancels with 1/2 from
$[k^{2}+q^{2}+(\k+\q)^{2}] \to 2 (1+q^{2}+q\cos\vartheta)$.

For $\eps=0$, the prefactor ${S_{d}^{2}}/{(2\pi)^{2d}}$ in (\ref{D2}) is
replaced with $1/64\pi^4$, while the angular averaging acquires the form
$$ \langle\langle \dots \rangle\rangle = \frac{2}{\pi}\, \int^{\pi}_{0}\,
\sin^{2} \vartheta\, \dots\,  . $$
Calculating the resulting double integral gives ${\cal R}= (3/2) \ln(4/3)$.

Now consider the total cofactor which contains the poles in $\eps$ and $y$.
It comes from the denominators in (\ref{D10}) and has the form:
\begin{eqnarray}
{\cal P} = \left\{ \frac{1}{2\eps}+ \frac{3 w^{2}}{2\eps-2(\eps-y)}
+\frac{3w}{2\eps-(\eps-y)} + \frac {w^{3}} {2\eps-3(\eps-y)} \right\}.
\label{D7}
\end{eqnarray}
Then constant $Z_{1}$ which cancels the poles in (\ref{Gamma}) will
have the form
\begin{eqnarray}
Z_{1}= 1 - \frac {\lambda^{2}}{6}\,   {\cal R} {\cal P}
\label{D8}
\end{eqnarray}
($2\sigma$, being an overall factor in the first term of (\ref{Gamma}) and
in (\ref{D1}), does not enter the expression for $Z_{1}$).
The anomalous dimension is
$$\gamma_{1} = \tilde{\cal D}_{\mu} \ln Z_{1}$$
in renormalized variables using the chain rule we obtain
($u=\lambda/16\pi^2$)
$$\gamma_{1} = [\beta_{u}\partial_{u}+\beta_{w}
\partial_{w}] \ln Z_{1}$$
with $\beta$ functions from (\ref{betas}). In our approximation,
$$ \tilde{\cal D}_{\mu}=   -\eps {\cal D}_{u} +
(-y+\eps) {\cal D}_{w} \quad {\rm and} \quad
\ln Z_{1} = - \frac {\lambda^{2}}{6}\,   {\cal R} {\cal P}. $$
This finally gives
\begin{eqnarray}
\gamma_{1} = \frac{\lambda^{2}}{6} \, {\cal R}(1+w)^{3} =
u^{2}(1+w)^{3}\, \ln(4/3),
\label{D9}
\end{eqnarray}
as already stated in (\ref{NL2}). For $w=0$ one obtains
$\gamma_{1} = u^{2}\ln(4/3)$, in agreement with the result known
for the model A; see e.g. \cite{Book3,AV}.

In the same manner we can derive the other results given in (\ref{NL2}).
For $\gamma_{2,3}$ the integral is
quadratic and it should be expanded in the external frequency and momentum
to $\Omega$ and $p^{2}$; the coefficients will be logarithmic integrals
of the type (\ref{D1}) and we can proceed as before for (\ref{D2}).
For $\gamma_{4,5}$ this is much simpler because the diagrams are one-loop
ones, they are logarithmic, there is only one momentum $\k$ and the
trick involving the differentiation ${\cal D}_{m}$ is not needed.

\end{document}